\documentclass[]{spie}  %>>> use for US letter paper
\usepackage[]{graphicx}

\title{Contact effects in polymer field-effect transistors} 

\author{D. Natelson\supit{a}\supit{b}, B.H. Hamadani\supit{a}, J.W. Ciszek\supit{c}, D.A. Corley\supit{c}, J.M. Tour\supit{c}\supit{d}
\skiplinehalf
\supit{a}Department of Physics and Astronomy, Rice University, Houston,
TX 77005
\supit{b}Department of
Electrical and Computer Engineering, Rice University, Houston,
TX 77005
\supit{c}Department of Chemistry, Rice University, Houston,
TX 77005
\supit{d}Department of Computer Science and the Smalley Institute for Nanoscale Science and Technology, Rice University, Houston,
TX 77005
}

\authorinfo{Further author information: (Send correspondence to D.N.)\\D.N.: E-mail: natelson@rice.edu, Telephone: 1 713 348 3214}
\begin{document} 
  \maketitle 

%%%%%%%%%%%%%%%%%%%%%%%%%%%%%%%%%%%%%%%%%%%%%%%%%%%%%%%%%%%%% 
\begin{abstract}
Contact resistances often contribute significantly to the overall device resistance in organic field-effect transistors (OFETs). Understanding charge injection at the metal-organic interface is critical to optimizing OFET device performance.  We have performed a series of experiments using bottom-contact poly(3-hexylthiophene) (P3HT) OFETs in the shallow channel limit to examine the injection process.  When contacts are ohmic we find that contact resistivity is inversely proportional to carrier mobility, consistent with diffusion-limited injection.  However, data from devices with other electrode materials indicate that this simple picture is inadequate to describe contacts with significant barriers.  A generalized transmission line method allows the analysis of nonohmic contacts, and we find reasonable agreement with a model for injection that accounts for the hopping nature of conduction in the polymer.  Variation of the (unintentional) dopant concentration in the P3HT can significantly alter the injection process via changes in metal-organic band alignment.  At very low doping levels, transport suggests the formation of a barrier at the Au/P3HT interface, while Pt/P3HT contacts remain ohmic with comparatively low resistance.  We recently observed that self-assembled monolayers on the metal source/drain electrodes can significantly decrease contact resistance and maintain ohmic conduction under conditions that would result in nonohmic, high resistance contacts to untreated electrodes.  Finally, we discuss measurements on extremely short channel devices, in the initial steps toward examining transport through individual polymer chains.
\end{abstract}

%>>>> Include a list of keywords after the abstract 


%%%%%%%%%%%%%%%%%%%%%%%%%%%%%%%%%%%%%%%%%%%%%%%%%%%%%%%%%%%%%
\section{INTRODUCTION}

Understanding and controlling charge injection and removal in organic
electronic devices is a problem of great interest, from the basic
physics of the metal/organic interface to the practical need for
optimized device performance.  Contact resistances in inorganic
semiconductor devices are typically minimized by strong local doping
of the contact regions.  The high local carrier concentration is
intended to reduce the depletion length sufficiently to thin any
Schottky barrier to effective transparency.  Local doping of organic
semiconductors (OSCs) in FETs and organic light-emitting diodes
(OLEDs) is very challenging, particularly in solution-processable
devices.

\begin{figure}
   \begin{center}
   \begin{tabular}{c}
   \includegraphics[height=7cm]{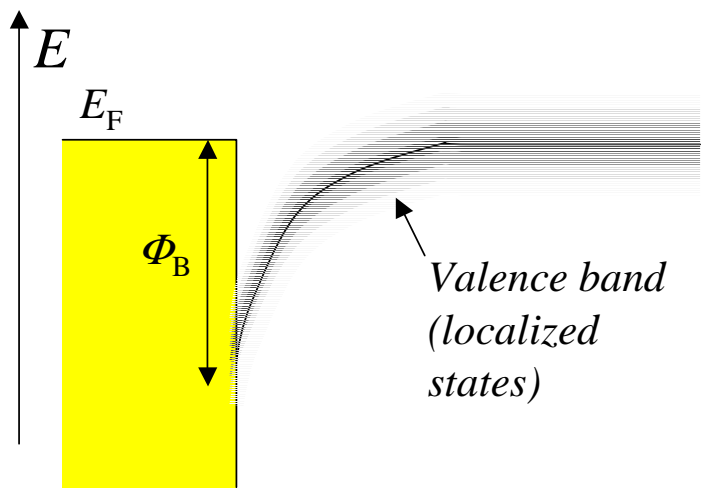}
   \end{tabular}
   \end{center}
   \caption[diagram] 
   { \label{fig:diagram} 
Diagram of band energetics at the metal/OSC interface.  The gradation in the OSC valence band represents the density of localized states as a function of energy, with a maximum around the band center.  The interfacial band offset as shown is a Schottky barrier for hole injection from the metal into the OSC.  The exact barrier height, interfacial charge transfer, and band bending self-consistently depend on the details of the OSC and the interface.  
}
   \end{figure}

Figure~\ref{fig:diagram} shows the basic situation, and makes clear
why this is a complicated problem.  Polymer OSCs are generally highly
disordered, and charge transport through these materials occurs via
variable range hopping through a strongly energy dependent density of
localized states\cite{PRB57Vissenberg}.  The detailed microstructure
of the OSC can drastically affect transport properties, with
increasing microscopic order generally correlating with higher charge
mobilities.  For the case of hole injection as shown, one can define a
Schottky barrier height as the energetic difference between the Fermi
level of the metal electrode and the center of the valence band of the
OSC.  This energetic alignment depends critically on the details of
the metal/OSC interface.  A detailed treatment of this problem in the
absence of significant metal/OSC charge transfer has been presented by
Arkhipov {\it et al.}\cite{PRB59Arkhipov}.  No complete theoretical
picture of this process has been developed.  In addition to the
metal/OSC interface, the OSC/dielectric interface is also of critical
importance in OFETs, thanks to its influence on the OSC microstructure
and proximity to the accumulated charge, which is typically confined
to a channel only a few nanometers thick adjacent to that surface.

%-------------

OFETs can be made in two configurations.  In top contact devices a
uniform layer of the OSC is deposited on the gate dielectric surface,
followed by source and drain electrodes, often via physical vapor
deposition of a metal.  The advantage of this approach is that the OSC
microstructure is uniform and well defined.  There are several
disadvantages.  Carriers must necessarily pass through some thickness
of undoped OSC to reach the channel region from the injecting contact,
and to reach the collecting contact from the channel.  The metal/OSC
interface is poorly controlled, since its formation generally involves
exposure of the organic layer to hot metal vapor.  Also, the
sensitivity of OSCs to organic solvents usually precludes lithographic
patterning of the source and drain electrodes.  Bottom contact
devices, in which the OSC is deposited on top of prepatterned source
and drain electrodes, avoid all three of these disadvantages.
However, the price paid is that the microstructure of the OSC on top
of the source and drain and at the electrode/dielectric interface can
differ significantly from that in the bulk of the channel.  

As OSC material quality and FET performance have improved, contact
issues are beginning to receive increased attention from the research
community.  Contact problems are generally far more severe in FET
structures, since current densities in technologically useful OFETs
can be several orders of magnitude higher than in OLEDs.  This paper
reviews our progress over the last several years in examining charge
injection and contact effects in OFETs.  Because of the enormous
parameter space of OSCs, electrode materials, and surface treatments,
we have limited our investigations to variations based on a single
conjugated polymer, P3HT, a single gate dielectric (SiO$_{2}$), and a
single OSC deposition process (drop casting).

\section{DEVICE FABRICATION AND CHARACTERIZATION}\label{sect:fab}

Except where discussed explicitly in Sect.~\ref{sect:interf}, all
devices fabricated for these experiments were made using identical
procedures.  The OFETs are in a bottom-contact configuration, using
degenerately doped $p+$ Si as the gate and substrate.  The gate
dielectric is 200~nm of thermally grown SiO$_{2}$.  Source and drain
electrodes are patterned using electron beam lithography and deposited
by electron beam evaporation in a vacuum of 10$^{-6}$~mB or better.
Following liftoff of the remaining resist, the resulting substrates
are cleaned for one minute in an oxygen plasma to remove possible
organic residue from the lithography process.  Plasma exposure times
are varied depending on the metal used for the source and drain
electrodes.  Device performance depends {\em critically} on the
cleaning of the dielectric and source/drain electrode interfaces.
Different cleaning procedures can strongly adversely affect measured
channel mobilities and contact resistances.

The P3HT is 98\% regioregular from Aldrich, as-received, prepared in a
solution (ranging from 0.02 to 0.06\% by weight) in chloroform that
has been passed through a 0.2~$\mu$m pore size polytetrafluoroethylene
(PTFE) filter.  The polymer is drop-cast by micropipette onto the
prepared surface, and the solvent is allowed to evaporate under
ambient conditions.  Typical film thicknesses are tens of nanometers,
as determined by atomic force microscopy.  Excess P3HT is removed from
larger electrode pads using a lint-free swab moistened with
chloroform.  Measured device characteristics do not correlate
obviously with P3HT thickness, except in terms of bulk conductivity
due to the presence of unintended dopants within the polymer.

\begin{figure}
   \begin{center}
   \begin{tabular}{c}
   \includegraphics[width=12cm]{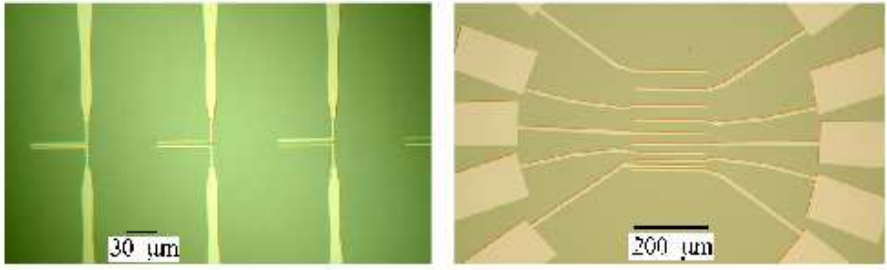}
   \end{tabular}
   \end{center}
   \caption[devices] 
   { \label{fig:devices} 
Interleaved Au source/drain electrodes patterned by e-beam lithography.  
The underlying substrate is the gate electrode.  Arrays of FETs
allow delineation between contact and bulk conduction phenomena.
}
   \end{figure}

Figure~\ref{fig:devices} shows an optical micrograph of a typical
array of interleaved electrodes prior to OSC deposition.  This
configuration allows sequential measurements of a series of devices
with a variety of channel lengths, all prepared and cast
simultaneously.

%-------------

The resulting structures are operated as standard accumulation-mode
FETs.  Measurements are performed in a variable temperature vacuum
probe station at a base pressure $\sim~10^{-6}$~mB, with movable
probes used to contact source, drain, and gate electrodes.  Device
characteristics are measured using a semiconductor parameter analyzer,
with the source electrode grounded.  

As deposited, the P3HT is moderately doped with holes.  This doping is
manifested by readily detectable bulk conduction between source and
drain at zero gate voltage, $V_{\rm G}=0$, or in two-terminal devices
with no gate at all.  The amount of this unintentional doping,
apparently due largely to atmospheric contamination, can be reduced
strongly by vacuum annealing at modest temperatures (330~K-370~K) for
a few hours.  Discussions below of doping (Sect.~\ref{sect:doping}) and
interfacial energetics (Sect.~\ref{sect:interf}) address the effects
of dopant concentration on both field-effect mobility and contact
resistances.

\section{OHMIC INJECTION} \label{sect:ohmic}

\begin{figure} 
\begin{center} 
\begin{tabular}{c}
\includegraphics[height=7cm]{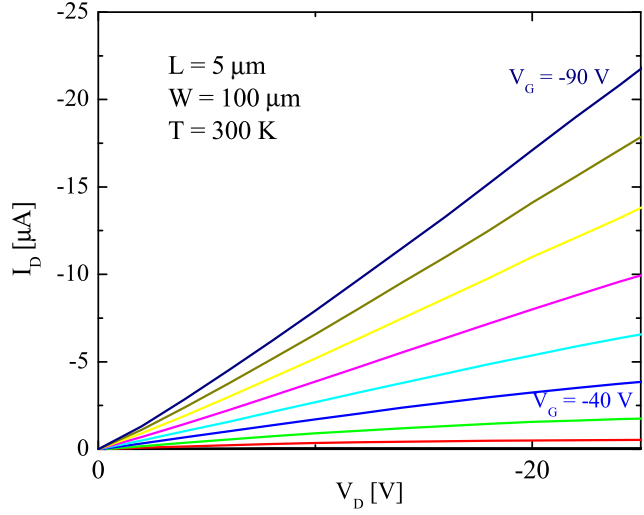} 
\end{tabular} 
\end{center}
\caption[ohmic]{ \label{fig:ohmic} Current-voltage characteristics
   at various gate voltages for a P3HT OFET with Au source/drain
   contacts, $W = 100~\mu$m, $L = 5~\mu$m, $T = 300$~K, after a few
   hours under vacuum.  Injection is ohmic.  } 
\end{figure}

In the presence of moderate doping (a few hours at room temperature in
vacuum after P3HT deposition) with clean Au source/drain electrodes,
and at all doping levels with Pt electrodes, source/drain conduction
at all gate voltages is found to be ohmic.  Figure~\ref{fig:ohmic}
shows typical $I_{\rm D}-V_{\rm D}$ traces at various $V_{\rm G}$ 
for such a device, far from saturation.

In the shallow channel ($V_{\rm D} < V_{\rm G}$) limit with ohmic
transport, it is comparatively simple to delineate between the contact
and channel contributions to overall device resistance using the
well-established transmission line method.  At a given $V_{\rm G}$,
the total device resistance, $R_{\rm on}$ may be measured for a series
of OFETs of fixed width, $W$, and varying channel length, $L$.
Plotting $R_{\rm on}$ vs. $L$ reveals a linear dependence, with the
intercept (extrapolating to $L=0$) giving $R_{\rm s}$, the series
parasitic resistance due to source and drain contacts.  The 
gate dependence of the slope allows the extraction of the
true field-effect mobility of the channel:
\begin{equation}
\frac{\partial\left[ \left(\frac{\partial R_{\rm on}}{\partial L}\right)^{-1}\right]}{\partial V_{\rm G}} = \mu(V_{G},T)W C_{\rm ox}.
\label{eq:mobility}
\end{equation}
Here $C_{\rm ox}$ is the capacitance per area of the gate oxide.

\begin{figure}
   \begin{center}
   \begin{tabular}{c}
   \includegraphics[height=7cm]{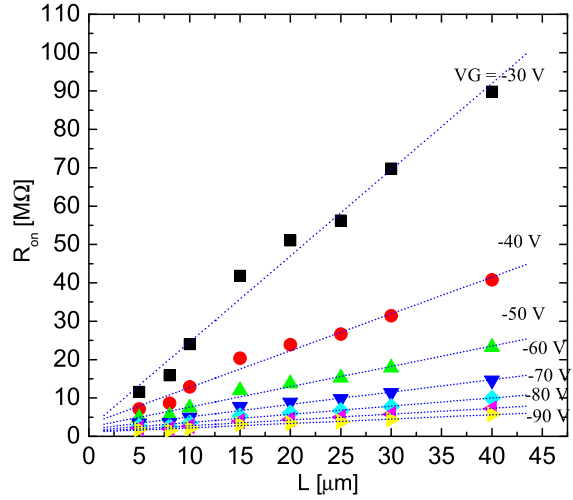}
   \end{tabular}
   \end{center}
   \caption[lengthscaling] 
   { \label{fig:lengthscaling} 
Total device resistance, $R_{\rm on}$, as a function of channel
length for a series of P3HT OFETs with Au source/drain contacts,
$W = 100~\mu$m, $T = 300$~K, after a few hours under vacuum.
The slope of the trendlines is proportional to $\mu$, 
and the $L = 0$ intercept is proportional to the contact resistivity. 
}
   \end{figure}

Figure~\ref{fig:lengthscaling} shows this kind of analysis on a series of
$W = 100~\mu$m devices at room temperature.  One challenge with this
method is that, for devices with relatively low contact resistances,
device-to-device variations in properties can lead to relatively large
uncertainties in $R_{\rm s}$.  

Scanning potentiometry
measurements\cite{APL78Seshadri,APL80Burgi,JAP94Burgi} allow direct
measurement of the voltage drop at the contact regions.  For the case
of ohmic injection, such measurements have shown that the contact
voltage drop, $V_{\rm c} \equiv I_{\rm D}R_{\rm s}$, occurs primarily
at the injecting contact, in this case the source.  For nonohmic 
contacts, the situation is more complicated, as discussed below.

   \begin{figure}
   \begin{center}
   \begin{tabular}{c}
   \includegraphics[height=7cm]{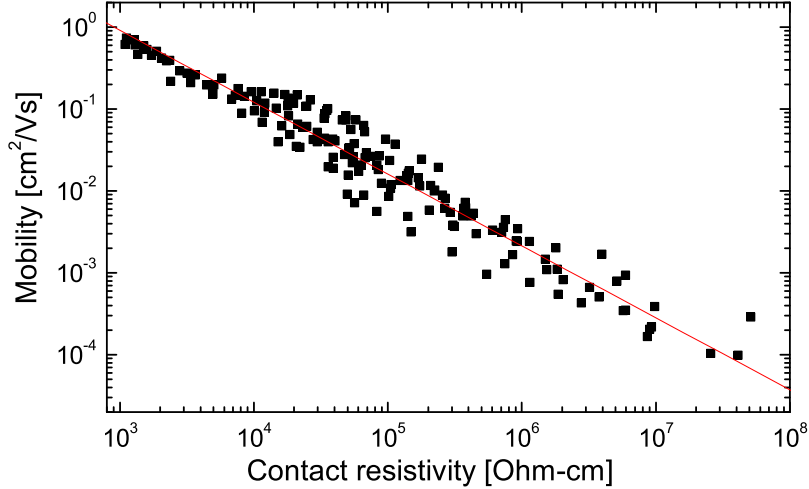}
   \end{tabular}
   \end{center}
   \caption[dli] 
   { \label{fig:dli} 
Mobility vs. contact resistivity measured in a large ensemble
of bottom-contact Au/P3HT FETs of widths 5, 30, and 100~$\mu$m, from 300~K
down to 100~K, with $V_{\rm G}$ from -10 to -90~V.  Adapted from \protect{\cite{HamadanietAl04APL}}.
}
   \end{figure}

Analyzing both $R_{\rm s}$ and $\mu$ as a function of
temperature down to $\sim$~100~K reveals an Arrhenius dependence for
both quantities, with very similar activation energies.  Indeed, a
direct comparison of contact resistivity and $\mu$ across
many devices, many gate voltages, and a broad temperature range
shows\cite{HamadanietAl04APL} that $R_{\rm s}W \propto \mu_{\rm
FE}^{-1.09}$ over four decades in both $R_{\rm s}W$ and $\mu_{\rm
FE}$.  Similar trends are seen for P3HT devices with platinum
contacts.

This is understandable in the context of diffusion-limited injection,
originally suggested in the context of amorphous silicon
FETs\cite{PRLEmtage}.  This model has received renewed attention
recently\cite{ChemPhysLett299Scott,PRL86Shen,JVacSciTechScott},
including modification to include barrier-lowering effects at high
biases.
\begin{equation}
J_{\rm INJ}= 4 \psi^{2} N_{0} e \mu E \exp(-\phi_{B}/k_{\rm B}T)\exp(f^{1/2}),
\label{eq:scott1}
\end{equation} 
where $\psi$ is a slowly varying function of electric field, $E$,
that approaches 1 in the low $E$ limit;
$N_{0}$ is the density of localized states in the OSC at the
metal/OSC interface and the Fermi level of the metal; $\phi_{B}$ is 
Schottky barrier; and the $f = e^3 E/[4 \pi \epsilon \epsilon_{0}(k_{\rm
B}T)^{2}]$ term is the barrier lowering term relevant at large $E$.  
This model has received support from experiments in 
two-terminal OSC-metal diodes\cite{PRL86Shen}.  In the
absence of a barrier, the contact resistivity is expected to be
inversely proportional to $\mu$.  Conversely, a large barrier
in this model would result in a large difference between the
temperature dependences of $R_{\rm s}W$ and $\mu^{-1}$.
The effective contact resistance even in the absence of a large
energetic barrier is due to the hopping nature of the OSC.  When a
carrier is injected into the OSC, there is a competition between
diffusion by hopping away from the interface, and the attractive
interaction between the carrier and its image charge in the metal.
The result is diffusion-limited injection.  

Note that the broad nature of the valence band's energy-dependent
density of localized states implies that ``perfect'' band alignment
($\phi_{\rm B} = 0$) is not essential for ohmic injection.  Rather,
the contact resistivity implied by Eq.~(\ref{eq:scott1}) depends on
the OSC density of localized states at the interface at the metal
Fermi energy.  One can conceive of two different metal/OSC interfacial
energetic alignments that would both give ohmic injection, but with
different values of $N_{0}$.  Only when the barrier height (as defined
above) significantly exceeds the width of the valence band would
nonohmic injection be expected to occur.

\section{NONOHMIC INJECTION}\label{sect:nonohmic}

The apparent success of Eq.~(\ref{eq:scott1}) in describing
ohmic injection in Au/P3HT devices\cite{HamadanietAl04APL} naturally
suggests considering other source/drain electrode materials.
Variations in metal work function, while not exclusively 
determining the energetic alignment at the interface, can be
used to try to achieve larger $\phi_{\rm B}$ values to
study nonlinear injection\cite{APL73Hill}.

\begin{figure}
   \begin{center}
   \begin{tabular}{c}
   \includegraphics[height=7cm]{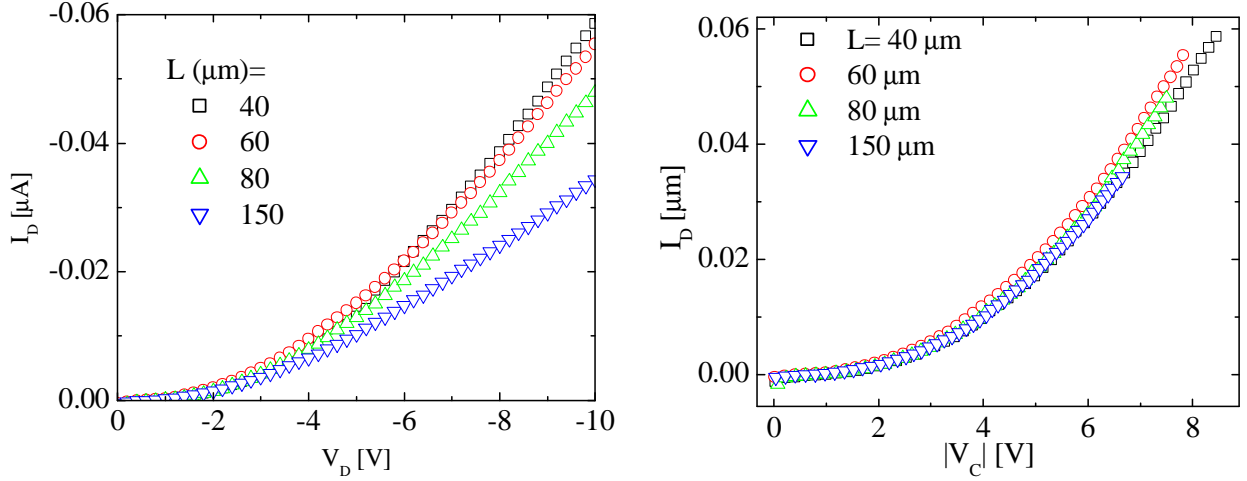}
   \end{tabular}
   \end{center}
   \caption[collapse] 
   { \label{fig:collapse} 
Left:  $I_{\rm D}-V_{\rm D}$ characteristics in the shallow channel regime
($V_{\rm G} = -70$~V, $T = 240$~K, $W = 400~\mu$m) for four copper source/drain
electrode P3HT FETs of different channel lengths.  Right:  The inferred
$I_{\rm D}-V_{\rm C}$ characteristics for the copper-P3HT contacts from
these devices, from Eq.~(\ref{eq:eq2}) with $\mu = 0.0038$~cm$^{2}$/Vs.
The collapse of all four device characteristics onto a single curve
strongly supports the conclusion that this procedure results in 
finding robust current-voltage characteristics for injecting contacts,
within its regime of validity.
}
   \end{figure}

Observing nonohmic injection is straightforward.  Using lower work
function metals such as copper, silver, and chromium, $I_{\rm
D}-V_{\rm D}$ characteristics like those in Fig.~\ref{fig:collapse} are
seen in devices made like those above.  The challenge is to
perform an analysis analogous to the transmission line method
for ohmic contacts, so that one can examine the injection process
specifically.  

We developed an approach to this problem based on the method
of Street and Salleo\cite{APLSalleo}.  Continuing to assume,
as seems borne out in scanning potentiometry\cite{APL78Seshadri,APL80Burgi,JAP94Burgi},
that most of the contact-induced potential drop occurs at
the injecting electrode, we again work in the shallow channel
regime and divide the total source/drain bias, $V_{\rm D}$, into
two components.
A contact voltage, $V_{\rm C}$, is assumed to drop across a
region of length $d$ near 
the injecting electrode, with the remaining $V_{\rm ch} = V_{\rm
D}-V_{\rm C}$ dropped across the main channel.  Using the charge control
model\cite{PhysSemicDevShur}, we write:
\begin{equation}
I_{\rm D} = W C_{\rm ox} \mu [V_{\rm G}-V_{\rm T}-V(x)]\frac{dV}{dx},
\label{eq:eq1}
\end{equation}
where $V(x)$ is the potential at some position $x$ in the channel and $V_{\rm
T}$ is the threshold voltage. Integrating Eq.~(\ref{eq:eq1}) over the channel without the contact region (from $x = 0$ to $L-d$) gives:
\begin{equation}
\frac{I_{\rm D}}{WC_{\rm ox} \mu}(L-d)=(V_{\rm G}-V_{\rm T})(V_{\rm D}-V_{\rm C})-\frac{1}{2}(V_{\rm D}^{2}-V_{\rm C}^{2}).
\label{eq:eq2}
\end{equation}

Given $I_{\rm D}$ vs. $V_{\rm D}$ for a particular device at a 
particular $V_{\rm G}$ and known $V_{\rm T}$ and $\mu$, Eq.~(\ref{eq:eq2}) 
can be used to infer $V_{\rm C}$ for each $I_{\rm D}$.
While $V_{\rm T}$ may be inferred for a given device from the
gate response at small $V_{\rm D}$, one needs additional 
information to find $\mu$, the true channel mobility.  This
is where the transmission line approach comes into play,
in which one can use dependence of $I_{\rm D}$ on channel
length to infer $\mu$.

At a given $T$ and $V_{\rm G}$, $I_{\rm D}-V_{\rm D}$ data is
collected from devices in such an array, and Eq.~(\ref{eq:eq2}) is
used with some assumed $\mu$ to infer corresponding $I_{\rm D}-V_{\rm
C}$ data for all the different channel lengths.  For a well controlled
fabrication process, the injection and bulk transport properties
should be the same in all the devices, implying that the correct value
of $\mu(V_{\rm G}, T)$ is the one for which analysis of each device
leads to {\em identical} $I_{\rm D}-V_{\rm C}$ characteristics.
A typical example of this is shown in Fig.~\ref{fig:collapse}.

Applying this approach to learn more about the nature of the nonohmic
injection process is described in some detail
elsewhere\cite{HamadanietAl05JAP}.  We find that the form and
particularly the temperature dependence of the $I_{\rm D}-V_{\rm C}$
characteristics for electrodes materials such as Cr and Cu are poorly
described by the nonzero $\phi_{\rm B}$ form of Eq.~(\ref{eq:scott1}).
These observations are consistent with the results of
others\cite{JAP94Burgi}.  

Instead we find that a more sophisticated
treatment\cite{JAP84Arkhipov,PRB59Arkhipov} of injection through a
barrier into a hopping conductor with a strongly energy dependent
density of localized states reasonably approximates the injection data
when ``sensible'' values of model parameters ({\it e.g.} the few
nm$^{-1}$ localization length$^{-1}$ in the P3HT; a barrier height of
0.2-0.3~eV) are assumed.  While this consistency suggests that this
treatment includes much of the relevant physics, it is wise to have
some concerns about the uniqueness of this description, given that the
data as a function of voltage and temperature are relatively smooth,
and the model contains several parameters that are difficult to assess
independently.  

Within this framework, the failure of the model of
Eq.~(\ref{eq:scott1}) comes from its underlying assumption of a single
$\mu$, reflecting a fixed density of localized hopping sites.  When
$\phi_{\rm B}$ is small and the electrode Fermi level is relatively
near the middle of the roughly Gaussian density of localized states of
the valence band, this approximation is not unreasonable.  However,
when $\phi_{\rm B}$ is larger and the initial hop from the metal into
the OSC occurs at an energy where the OSC density of localized states
is strongly energy dependent, detailed variable range hopping physics
comes into play\cite{PRB59Arkhipov,HamadanietAl05JAP}, lowering the
effective barrier.

One important parameter from that analysis is an estimate of the
depletion distance $d$ over which $V_{\rm C}$ is dropped of around
100-200~nm.  Note that prior to these measurements $d$ was already
constrained in both large and small limits.  For $d < \sim~10$~nm, the
internal electric fields $\sim V_{\rm C}/d$ at the contact region
would have exceeded the breakdown field of sensible materials.  In the
other limit, if $d$ was significantly larger than 200~nm, scanning
potentiometry measurements would resolve the potential drop near the
contact.  The physical origin of this length scale should be
intimately related to the detailed physical mechanism for injection.
As alluded to above, no current theoretical treatment exists that
includes all the physically interfacial charge transfer between the
metal and the OSC, a likely component necessary for a description of
an intrinsic depletion effect.

\section{DOPING AND INJECTION}\label{sect:doping}

\begin{figure}
   \begin{center}
   \begin{tabular}{c}
   \includegraphics[height=7cm]{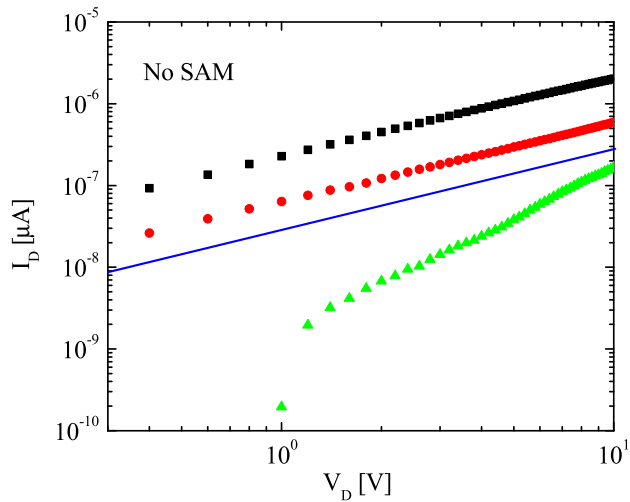}
   \end{tabular}
   \end{center}
   \caption[anneal] 
   { \label{fig:anneal} 
Sets of $I_{\rm D}-V_{\rm D}$ data for an $L = 40~\mu$m, $W = 200~\mu$m 
Au/P3HT OFET at 300~K and $V_{\rm G} = -70$~V.  Top to bottom,
curves are subsequent annealing cycles in vacuum (top: 16~h
at 300~K; middle:  additional 18~h at 350~K; bottom:  additional  
18~h at 360~K).  Adapted from Hamadani {\it et al.}\protect{\cite{HamadanietAl06NL}}.
The solid line shows ohmic transport.}
   \end{figure}

Further experiments on nonohmic injection were prompted by
observations of increasingly nonlinear $I_{\rm D}-V_{\rm D}$
characteristics in Au/P3HT devices upon further annealing in vacuum.
An example of this is shown in Fig.~\ref{fig:anneal}.  Each annealing
step corresponds to at least 12 hours at temperatures between 330~K
and 350~K.  This effect was completely reversible upon a brief (few
hours) exposure to ambient air at room temperature.  The general
chemical inertness of Au, the temperature range, and the reversibility
suggest strongly that this effect is not the result of interfacial
chemistry between the P3HT and the electrodes.  

The primary effect of the annealing process appears to be a reduction
in the concentration of hole-like carriers at $V_{\rm G}=0$ as
manifested in the two-terminal source/drain conductivity.  These
carriers are associated with some kind of unintended chemical doping
upon exposure to ambient air.  The precise nature of the dopants and
doping mechanism remains under investigation\cite{JAP95Hoshino}.  This
decrease in doping is accompanied by a corresponding decrease in
field-effect mobility.  In light of the observed increase of $\mu$
with increasing accumulated charge density, this correlation is
unsurprising.  Unfortunately it is not easy to deconvolve the carrier
concentration from the bulk mobility at $V_{\rm G}=0$ to obtain the
doping density as a function of annealing conditions and time.

The nonohmic conduction, also observed by others\cite{Rep2003} in
two-terminal Au/P3HT devices, is clearly a contact effect.  All other
things being equal, for a given annealing schedule, shorter channel
devices become more nonohmic than longer channel devices.
Furthermore, the nonlinearity is more severe at larger magnitudes of
$V_{\rm G}$, when the bulk conduction in the channel is improved due
to the dependence of $\mu(V_{\rm G})$.  Once the $I_{\rm D}-V_{\rm D}$
nonlinearity is moderately severe, the scaling analysis of the
previous section fails, suggesting that one of its underlying
assumptions is no longer applicable.  The most likely candidate
explanation is that there is a significant interfacial voltage drop at
both injecting and collecting contacts.  This could be tested with
{\it in-vacuo} scanning potentiometry measurements.

Analogous annealing experiments in devices with Pt contacts, while
exhibiting the same drops in bulk conduction and similar drops in
$\mu$, show ohmic conduction under accumulation at annealing histories
far more severe than those shown to give strong nonlinearities in
Au/P3HT devices.  This naturally raises the question of what
distinguishes the Au/P3HT interface from the Pt/P3HT interface.  Given
that both metals are vacuum deposited in a similar manner and exposed
to the same processing conditions prior to and during P3HT deposition,
it seems likely that the difference in contact properties results from
some difference in the metals themselves.  One possible difference is
in work function.  Clean Au in ultrahigh vacuum (UHV) has a work
function of approximately 5.1~eV, while Pt under similar conditions is
closer to 5.6~eV.  Exposure to ambient processing conditions can
modify these values significantly, however.

Interface-sensitive spectroscopies are the ideal tools for examining
the level alignment energetics between metals and OSCs.  Studies
involving ultraviolet photoemission spectroscopy (UPS) can provide
information about filled electronic states, while inverse
photoemission can reveal much about empty states.  Unfortunately,
optimal results from these techniques require ultrahigh vacuum sample
preparation and extremely thin organic layers, two conditions rather
far removed from the actual devices in our experiments.  

However, through collaboration with the Rochester
group\cite{HamadanietAl05PRB}, we were able to learn important
energetic information about these material systems.  Under annealing
schedules similar to those described above, the valence band of the
bulk P3HT is observed to shift farther from the vacuum level.  This
shift is $\sim$ 500~meV for Au electrodes, but only 2-300~meV for Pt
electrodes.  This is qualitatively consistent with the observed
transport data, suggesting a shift of the transport level farther out
of the valence band density of localized states for Au than for Pt
under annealing.  The thickness of the sample films, comparable
to those used in the FET experiments, made it not possible to 
get direct information about the metal/OSC interface, however.

\section{CONTROLLING INTERFACIAL ENERGETICS}\label{sect:interf}

\begin{figure}
   \begin{center}
   \begin{tabular}{c}
   \includegraphics[height=7cm]{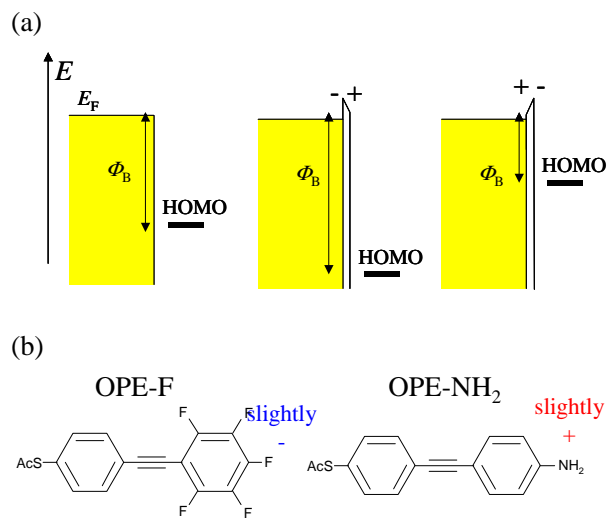}
   \end{tabular}
   \end{center}
   \caption[wkfn] 
   { \label{fig:wkfn} 
(a) A cartoon illustrating the naively expected results of modifying 
the source/drain electrode metal surface with an engineered interfacial
dipole of either sign (after de Boer {\it et al.}\protect{\cite{deBoer2005}}).  (b)
Two molecules used in our experiments\protect{\cite{HamadanietAl06NL}}
to increase (decrease) the effective Au work function, with the
intent of reducing (increasing) any injection barrier for holes from
Au.  
}
   \end{figure}

To test whether the difference in these doping effects lies in the
metal work function, perhaps through some kind of Fermi level pinning
at the metal/OSC interface, surface chemistry was used to engineer the
metal work function.  As shown in Fig.~\ref{fig:wkfn}a, a surface
dipole layer on the metal can alter the effective work function of the
metal, and therefore naively the band alignment at the interface.
This idea was first demonstrated in OLED devices over 10 years
ago\cite{Campbell1996}, and has received much attention from that
community\cite{Nuesch1998,Zuppiroli1999,Tour2001,deBoer2005}, with
comparably little work done in FET
structures\cite{Gundlach2001,Kim2003}.

Figure~\ref{fig:wkfn}b shows two types of molecules used to modify
the effective work function of the Au surface.  These molecules were
designed to self-assemble into ordered monolayers on the source and
drain electrodes via standard Au-thiol deprotection
chemistry\cite{Cai2002}.  Molecular design attempts to optimize the
combination of short, conjugated oligomers (intended to be relatively
electronically transparent) with terminal groups to impose a permanent
molecular dipole moment of appropriate sign.

As we reported recently\cite{HamadanietAl06NL}, self-assembly of the
molecules does, indeed, modify the energetics of the Au surface.
Comparative scanning potentiometry (same tip, all samples mounted
simultaneously) was performed under ambient conditions on SAM-modified
and non-SAM treated Au films prepared and cleaned as described in
Sect.~\ref{sect:fab}.  Assembly of the OPE-F molecule appears to raise
the work function (relative to the no-SAM Au) by as much as 0.9~eV,
and this change is stable in air for hours at least.  By contrast,
assembly of the OPE-NH$_{2}$ compound decreased the effective work
function by approximately 0.2-0.3~eV, though these SAM-treated
surfaces approached the same surface potential as unassembled surfaces
on the hour time scale.

\begin{figure}
   \begin{center}
   \begin{tabular}{c}
   \includegraphics[height=7cm]{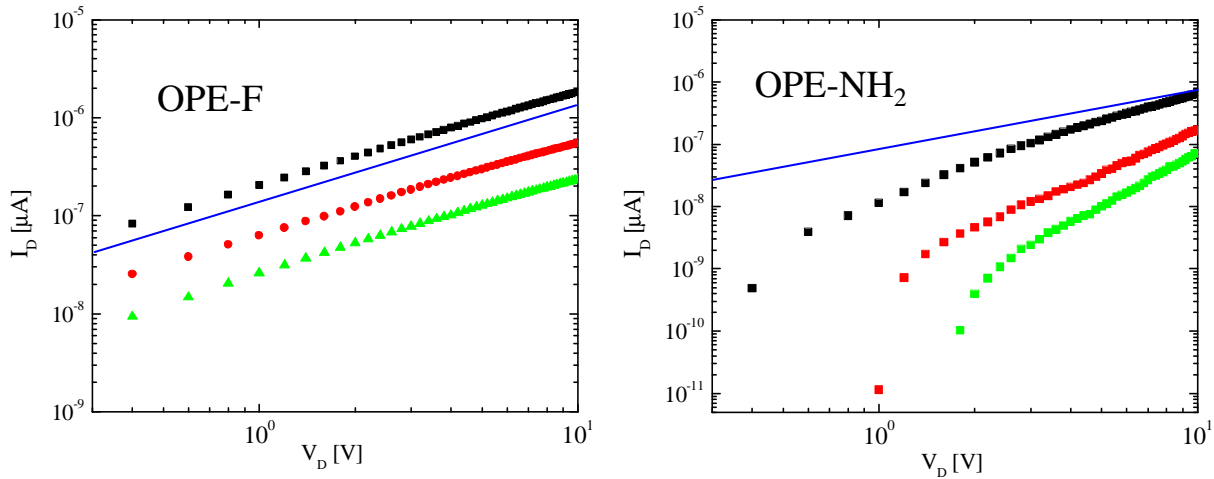}
   \end{tabular}
   \end{center}
   \caption[samresults] 
   { \label{fig:samresults} 
Left:  Transport through an OPE-F treated Au/P3HT device,
with a similar annealing schedule as that in Fig.~\ref{fig:anneal}.
ohmic transport (solid line) persists even at long anneal times.
Right:  Transport through an OPE-NH$_{2}$ treated Au/P3HT
device, demonstrating consistent, severe nonohmic injection,
similar to that seen in devices with Cr electrodes.  These
results are consistent with the trends expected if the 
SAM treatments are modifying the effective band alignment 
between the metal and the P3HT.  Adapted from Hamadani {\it et al.}\protect{\cite{HamadanietAl06NL}}.
}
   \end{figure}

Transport measurements are consistent with these results.  Devices
with OPE-F monolayers assembled on Au electrodes exhibit the same kind
of ohmic conduction seen in the Pt devices described above, even upon
relatively severe annealing.  Similarly, OFETs with OPE-NH$_{2}$
monolayers on the Au source and drain electrodes demonstrate nonlinear
injection even at moderate doping concentrations, similar to the
results seen in Cr electrode devices.  This is shown in
Fig.~\ref{fig:samresults}.  The nonlinear injection in the
OPE-NH$_{2}$ devices persists even after hours of exposure to ambient
conditions, suggesting that the instability seen in surface potential
of the bare SAM-coated film is somehow mitigated by the OSC overlayer.

These observations are encouraging, and demonstrate that surface
chemistry may be used to optimize the injection process (reducing
contact resistances, maintaining ohmic injection at low doping).
Future efforts to investigate the detailed energetic configuration of
the buried metal/SAM/OSC interface would be extremely useful,
particularly to determine to what extent the simple picture in
Fig.~\ref{fig:wkfn}a is accurate.  An eventual goal of this kind of
surface engineering would be the development of complementary OSC
devices.  Contacts to $n$-channel OFETs and cathodes in OLEDs often
involve the deposition of low work function materials ({\it e.g.} Al,
Ca) that tend to be extremely reactive and therefore inconveniently
unstable under ambient conditions.  A simple, stable surface treatment
might one day enable a single electrode metal to be used for
contacting either $p$ or $n$-channel devices.

\section{NANOSCALE DEVICES}\label{sect:nano}

The ability to fabricate extremely short channel devices combined with
our improved understanding of injection leads us to consider using
sub-100~nm channel OFETs as tools for further investigations of
injection and conduction.  A natural set of experiments suggested by
the results of Sect.~\ref{sect:nonohmic} is to fabricate OFETs with
channel lengths comparable to the inferred contact length scale $d$.
The responsiveness or lack thereof of the conduction in such devices
to $V_{\rm G}$, for example, should be one possible way of assessing
to what extent the transport properties in the channel are controlled
by interfacial charge transfer with the metal rather than by the gate
potential.  

Of particular interest would be OFETs with channel lengths smaller
than 10-30~nm, the typical size of a nanocrystalline ordered region
within P3HT\cite{APL69Bao,Nat404Sirring}, and close to the length of a
single polymer chain.  If devices could be constructed with favorable
energetic alignment at the metal/OSC interface, and with the entire
channel being ordered rather than a glassy hopping landscale, device
performance would give experimental access to unknowns such as the
ultimate limit of mobility in such materials.

   \begin{figure}
   \begin{center}
   \begin{tabular}{c}
   \includegraphics[height=7cm]{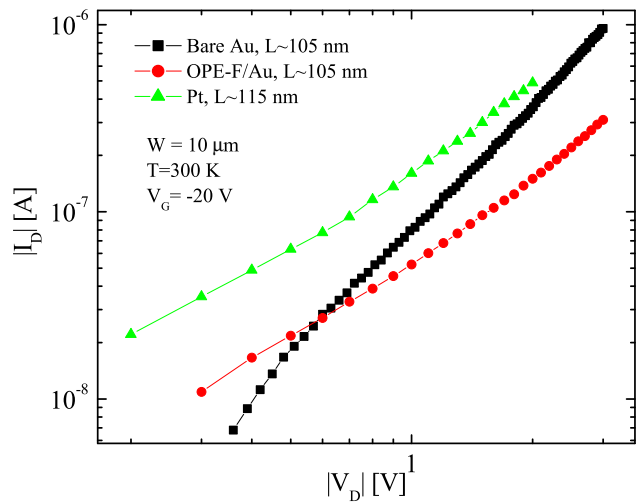}
   \end{tabular}
   \end{center}
   \caption[nano1] 
   { \label{fig:nano1} 
Transport through $\sim$~100~nm channel length devices with 
various electrode materials and treatments.  Pt and OPE-F
treated electrodes are expected to be in the bulk-limited
regime, and show ohmic conduction at low biases.  The
untreated Au electrodes show nonlinearities suggesting
contact limited conduction as described in the text.
}
   \end{figure}

Figure~\ref{fig:nano1} shows an interesting comparison of devices with
channel lengths $L \sim$~100~nm, $W = 10~\mu$m.  These OFETs have been
prepared with a two-stage e-beam lithography process with surfaces
cleaned as before, with no annealing beyond 2 hours in vacuum.  The
Au-only device is highly nonlinear already.  The devices based on
OPE-F SAM-treated Au electrodes and Pt electrodes, however, have
nearly identical functional forms of $I_{\rm D}$ vs.  $V_{\rm D}$,
both starting with an ohmic region extending to an average
source/drain electric field approaching $10^{7}$~V/m.  Previously
measured contact resistivities in larger devices made from the same
electrode materials suggest that these devices are contact limited,
and like those in Sect.~\ref{sect:ohmic} are manifesting ohmic contact
resistances.  Because of the diffusion-limited injection picture it is
still possible that the slight superlinear dependence of $I_{\rm
D}(V_{\rm D})$ in those two devices is a manifestation of the
intrinsic field-dependence of $\mu$ (see Hamadani and Natelson
(2004)\cite{HamadanietAl04JAP} and references therein).

These measurements are preliminary.  Experiments are ongoing with
shorter channel lengths and Al gates with native Al$_{2}$O$_{3}$ gate
dielectrics for improved gate coupling.  Other techniques including
electromigration\cite{ParketAl99APL} are under consideration for
enabling single-molecule level measurements.  Using such deeply scaled
devices as tools to understand basic device physics is just beginning.

\section{CONCLUSIONS}\label{sect:conclusions}

Injection and contact effects in OFETs remain rich and important
topics of investigation.  As described above, even restricting
experiments to a small part of the possible parameter space of
electrode materials, device geometries, OSC materials, and surface
preparations has revealed a wealth of information.  Simple pictures of
injection based on inorganic devices appear to be inadequate to
explain the observed features of polymer OFETs in the general case of
a significant energetic mismatch between the electrode Fermi level and
the OSC electronic structure.  Under certain circumstances it is
possible to extract the current/voltage characteristics of the
metal/OSC interface, and the resulting data constrain models of the
injection process.  Surface-sensitive spectroscopies are essential
tools for understanding the energetics of the metal/OSC interface, and
there is significant indirect evidence that electrode work function
and (unintentional) OSC doping level can strongly affect the injection
mechanism.  Surface chemistry provides a means of engineering
desirable injection properties through the use of dipolar SAMs to
alter the metal energetics.  Finally, sub-100~nm channel length OFETs
are going to be very useful tools for further investigations of basic
device physics questions.  

Much remains to be learned about the interfacial physics and chemistry
issues relevant to these problems.  Future developments in theory and
experimental tools for accessing buried interfaces prepared under
realistic (non-ideal) processing conditions will help greatly in the
optimization of resulting organic electronic and optoelectronic
devices.

%-------------

\acknowledgments     %>>>> equivalent to \section*{ACKNOWLEDGMENTS}       
 
D.N. acknowledges support from the David and Lucille Packard
Foundation, the Alfred P. Sloan Foundation, the Robert A. Welch
Foundation, the Research Corporation, and NSF grant ECS-0601303.
J.M.T. acknowledges support from DARPA and AFOSR.  D.N. also thanks
H. Ding and Prof. Y. Gao of the University of Rochester for useful
discussions and collaboration.

%%%%%%%%%%%%%%%%%%%%%%%%%%%%%%%%%%%%%%%%%%%%%%%%%%%%%%%%%%%%%
%%%%% References %%%%%

\end{document}